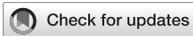

*CORRESPONDENCE
L. Olano-Vegas,
✉ l.olano@nanogune.eu
D. González-Díaz,
✉ diegogon@cern.ch

# Development of $Fe_2O_3$/YSZ ceramic plates for cryogenic operation of resistive-protected gaseous detectors

L. Olano-Vegas[1]*, I. Pardo[1], S. Leardini[1], M. Morales[1],
A. R. Carreira[2], R. M. Corral[2], D. González-Díaz[1]*, A. Tesi[3],
L. Moleri[3], C. D. R. Azevedo[4], L. Carramate[4] and F. Guitián[2]

[1]Instituto Galego de Física de Altas Enerxías (IGFAE), Universidade de Santiago de Compostela, Santiago de Compostela, Spain, [2]Instituto de Materiales, Universidade de Santiago de Compostela, Santiago de Compostela, Spain, [3]Department of Particle Physics and Astrophysics, Weizmann Institute of Science, Rehovot, Israel, [4]Institute of Nanostructures, Nanomodelling and Nanofabrication (i3N), Universidade de Aveiro, Aveiro, Portugal

We present a ceramic material based on hematite ($Fe_2O_3$) and zirconia stabilized with yttria at 8% molar (YSZ), that exhibits stable electrical properties with transported charge and that can be tuned to the resistivities necessary to induce spark-quenching in gaseous detectors ($\rho = 10^9\text{-}10^{12}$ $\Omega\cdot$cm), from room temperature down to the liquid-vapor coexistence point of nitrogen (77 K). It, thus, allows covering the operating temperatures of most immediate interest to gaseous instrumentation. The ceramics have been produced in a region of mass concentrations far from what has been usually explored in literature: optimal characteristics are achieved for $Fe_2O_3$ concentrations of 75%wt (LAr boiling temperature), 35%wt (LXe boiling temperature), and 100%wt (room temperature). The nine-orders-of-magnitude enhancement observed for the electrical conductivity of the mixed phases relative to that of pure $Fe_2O_3$ is startling, however it can be qualitatively understood based on existing literature. Plates of 4 cm × 4 cm have been manufactured and, prior to this work, operated in-detector at the LXe boiling point (165 K), demonstrating spark-free operation. Illustrative results obtained for the first time on a spark-protected amplification structure (RP-WELL) at around the LAr boiling point (90 K) are now presented, too.

# 1 Introduction

Resistive-protection schemes are nowadays ubiquitous in avalanche-based gaseous detectors for which the detection of single-photons (Melai et al., 2011) or single-electrons (Ligtenberg et al., 2018) is required, in harsh environments with an abundance of highly ionizing particles (Chefdeville et al., 2021) or when a very fast response is needed (Fonte et al., 2000; Lisowska et al., 2023). During detector operation, when the regular avalanche growth unnaturally destabilizes into the formation of a streamer channel, resistive-protection effectively helps at quenching further evolution and preventing electrical breakdown. At much lower gains than the ones needed for the aforementioned applications, operation in ultra-pure noble gas can in principle benefit from a similar protection scheme (Leardini et al., 2023), thus mitigating the impact of the feedback processes that eventually lead to discharge in those conditions.

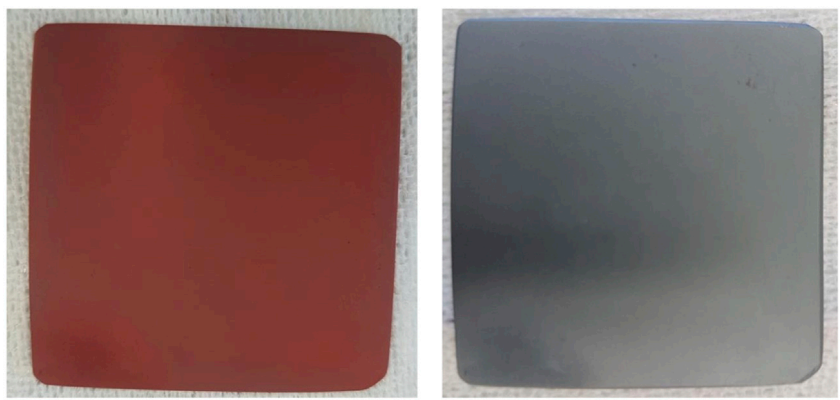

FIGURE 1
Left: pre-sintered ceramic plate (after polishing and edge-rounding) made from $Fe_2O_3$/YSZ (mass fraction 85/15), at the Instituto de Cerámica de Galicia (ICG/i-MATUS) in Spain. Right: fully-sintered ceramics plate (4 cm × 4 cm × 0.2 cm).

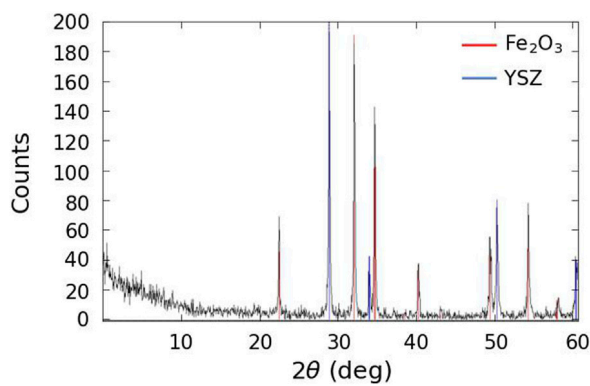

FIGURE 2
Results from X-ray dispersion on $Fe_2O_3$/YSZ ceramics (mass fraction 75/25), showing the main $Fe_2O_3$ and YSZ peaks.

When it comes to spark quenching through resistive protection, three main approaches exist: coatings (Lv et al., 2020), micro-patterned resistive elements (Chefdeville et al., 2021), and plates (Deppner, 2012). Although the first two represent attractive directions, the development of plates remains crucial in some detector configurations, chiefly for resistive plate chambers (RPCs) (Parkhomchuck et al., 1971; Santonico and Cardarelli, 1981) and resistive plate wells (RP-WELLs) (Jash et al., 2022). The scarcity of materials suitable for use as plates was a concern in the early 00s (e.g. Morales et al., 2013)), however the situation has been greatly alleviated over the last 2 decades with the introduction of 'chinese' glass (Wang, 2013), warm-glass (González-Díaz et al., 2005) and $Si_3N_4$/SiC ceramics (Laso García et al., 2016), an effort largely driven by the R&D of the Compressed Baryonic Matter experiment (CBM) at FAIR (Darmstadt) (Deppner, 2012). More recently, other materials have become available too (see for instance (Liu et al., 2020) and references therein). Stimulated by this progress, a natural follow-up question to us was whether resistive-plate protection of gaseous detectors could be extended to temperatures considerably below standard conditions.

It must be noted that the region of greatest interest to modern gaseous instrumentation starts at 87.5 K (the boiling point of argon) and new material technologies should cover at least the 165 K landmark (the boiling point of xenon), ideally up to room temperature. The difficulty of spanning such a broad temperature range lies in the fact that many insulator materials display an Arrhenius-type conductivity law: a material with the right resistivity for spark quenching at around 77 K can be expected to be a decent conductor at room temperature. Studies performed in the context of resistive plate chambers point to the need of volume resistivities around $10^9$–$10^{12}$ $\Omega$·cm to achieve good spark-quenching (e.g. (Francke et al., 2003)). Further, the material should ideally display a stable conductivity during operation as a function of time and transported charge, not being this the case for most common insulators (Morales et al., 2013). Therefore, and given that the detector atmosphere should be driven by the detector operational characteristics and not by the need to replenish the ionic charge carriers that migrate through the resistive protection, electron conductivity is an attractive feature.

When targeting a new insulating material focused on this latter aspect, ceramics represent an appealing possibility due to the large margin offered for customization, the possibility of achieving low outgassing and ruggedness, and their potential for scalability to large areas at affordable price. $Fe_2O_3$ (hematite), in particular, represents an interesting starting point as it displays electron conductivity based on the polaron-hopping mechanism (Liao et al., 2011; Ahart et al., 2022). Despite being too resistive for cryogenic use in gaseous detectors, its conductivity might be enhanced by orders of magnitude through the addition of Zr, Ti, Si or Ge (Morin, 1951; Kennedy et al., 1981; Ce et al., 2006; Kumar et al., 2011). In this case the dopant acts as an electron donor, up to concentrations of 10% Zr atoms per Fe atom (or 16.3%wt, in weight percentage), when a new phase is formed. For larger concentrations, $ZrO_2$ (zirconia) might be used instead, leading to speculation of additional contributions to the electron conductivity, either through the grain-boundary (Boukamp, 2003), or electron hopping through Fe cations in the zirconia lattice (Kharton, 1999). Given this landscape, a natural possibility was to evaluate ceramics based on $Fe_2O_3$ and zirconia stabilized with yttria at 8% molar (hereafter yttria-stabilized zirconia, YSZ) for which a large body of experience exists ((Verkerk et al., 1982; Slilaty, 1996; Gao et al., 2008; Molin et al., 2009; Bohnke et al., 2014; Guo, 2021) and references therein).

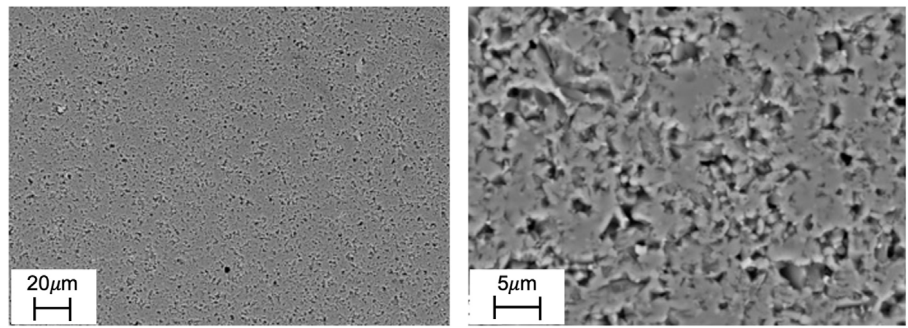

FIGURE 3
SEM images from $Fe_2O_3$/YSZ ceramics (mass fraction 75/25), at 20 μm (left) and 5 μm (right) scales.

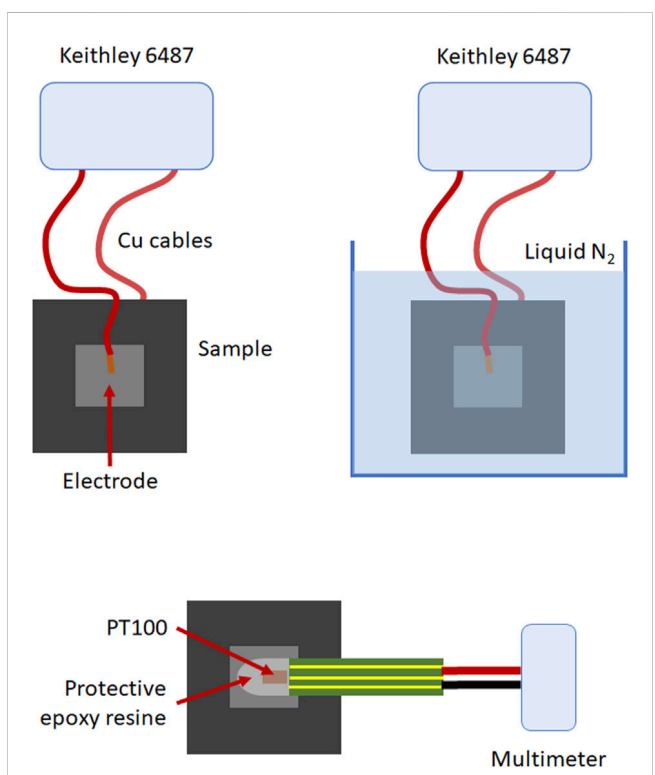

FIGURE 4
Description of the measurements of the electrical resistivity: electrodes were deposited on both faces of the sample, electrical leads clamped, and the readout performed through a Keithley 6487 picoammeter. Top: configuration for RT (left) and LNT (right). Bottom: PT-100 epoxied to the sample for resistivity measurements as a function of temperature. Temperature measurements were directly recorded with a multimeter.

Despite the existence of abundant experimental work, we found ourselves quickly in the range above the solubility limit of $Fe_2O_3$ in YSZ (about 6% molar (Molin et al., 2009)), that has been much less studied in literature.

In this work we evaluate the viability of using ceramics based on hematite ($Fe_2O_3$) and yttria-stabilized zirconia (YSZ) as resistive plates for spark-quenching of gaseous detectors, with $Fe_2O_3$ concentrations in the range of 30%–100% per weight (%wt). This paper is structured as follows: in Section 2 we describe the fabrication procedure, Section 3 introduces the setup and experimental methods; Section 4 includes a detailed characterization as a function of electric field, temperature, frequency, transported charge and $Fe_2O_3$ concentration; Section 5 presents first results from an RP-WELL operated at LArT, and we end with a short discussion in Section 6 and main conclusions in Section 7.

## 2 Fabrication

Ceramic plates were produced through slip casting. YSZ with crystallite sizes of 22 nm, and hematite of grain size below 5 μm (purity above 99%) were employed. Raw materials were weighted on a precision scale and admixed with 25% water and 1% deflocculant (Dolapix PC67). The resulting slurry was stirred in a mill filled with zircon balls for 24 h, after which it was poured into a plaster mold of internal dimensions 50 × 50 × 3 $mm^3$. It was then put on a vibrating table for some minutes and, once the sample was consolidated, it was dried in air.

The sintering process started with the sample heated in an oven up to 900°C, in air atmosphere (pre-sintering). After the sample was further consolidated and contracted down to about the final dimensions of 40 mm × 40 mm, it was thinned with SiC sandpaper down to 1–3 mm to ensure plano-parallelism. The final sintering was accomplished at 1250°C–1300°C for about 2 h, again in air atmosphere, and the sample let cool down at about 2 °C/min. Although magnetite formation ($Fe_3O_4$) can be expected only above 1345°C, a magnetic hysteresis analysis confirmed its absence from the final samples.

Finally, the external oxide layer stemming from the sintering process was removed by thinning the samples further down by about

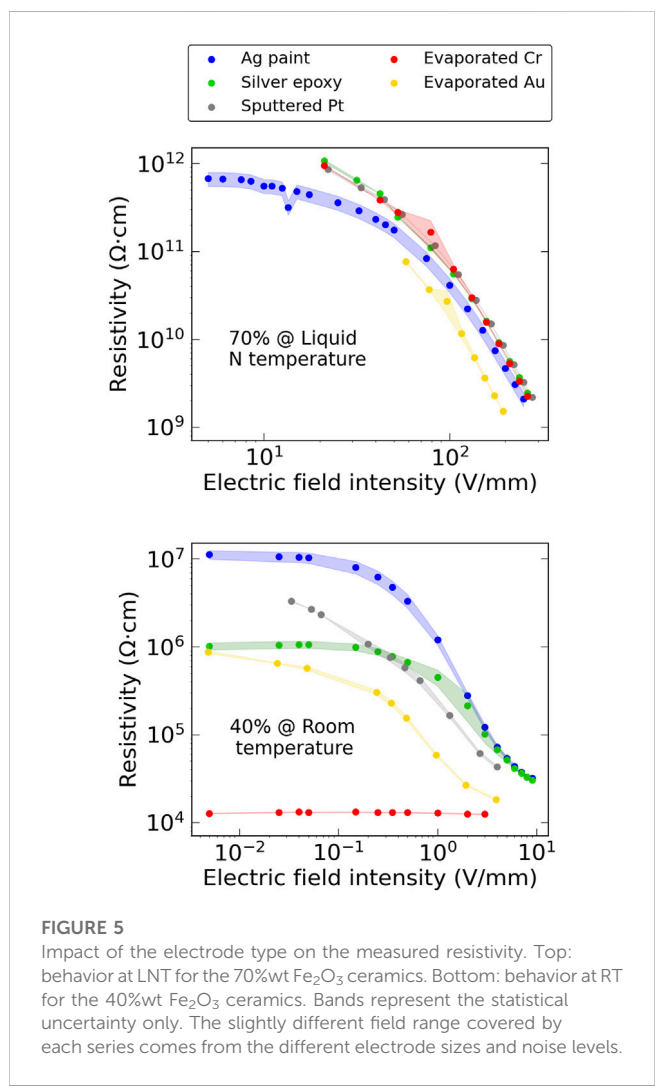

FIGURE 5
Impact of the electrode type on the measured resistivity. Top: behavior at LNT for the 70%wt $Fe_2O_3$ ceramics. Bottom: behavior at RT for the 40%wt $Fe_2O_3$ ceramics. Bands represent the statistical uncertainty only. The slightly different field range covered by each series comes from the different electrode sizes and noise levels.

100 μm. The criteria employed was that the surface electrical resistance would remain constant upon further polishing. Ceramic pieces are shown for illustration in Figure 1-left (pre-sintered) and right (sintered).

Samples were studied by X-ray dispersion and found to be largely free from contaminating species (Figure 2). The results from the diffractogram were generally found to be consistent with the mass fraction targeted during the production of the slurry. Surface analysis showed good homogeneity, compatible with the grain sizes from the raw materials, without strong indications of clustering (Figure 3). A study of the characteristic X-ray emission was compatible with the absence of species other than Fe, Zr, Y, and O. Further details on the fabrication process and morphological characterization will be given in forthcoming publications.

# 3 Experimental setup for electrical characterization

For plate-based resistive protection of gaseous detectors, the product of the bulk resistivity times the sample thickness represents the main metric driving system performance (e.g. (Deppner, 2012)). Accordingly, a thorough characterization of the bulk resistivity was performed following the recommendations in (IEC 62631-3-1, 2016). Guard-ring was dimmed necessary only for low-temperature measurements of the ceramics with low $Fe_2O_3$ concentration (30% wt), otherwise measurements would be shunted by surface currents. Figure 4 describes the setup. The characteristic size of the electrodes was chosen to be at least 10 times larger than the plate thickness (1–3 mm, hence electrodes were in the range 1–10 $cm^2$), to keep fringe-field effects at the 10% level or below (Lisowski and Kacprzyk, 2006). Connections were made through electrical leads clamped to the sample, that was immersed directly in a liquid $N_2$ (LN) bath in the case of measurements performed at LN-temperature (LNT). Different electrodes were studied: Ag-loaded epoxy (CHEMTRONICS, 2023), Ag-loaded paint (SAFETY DATA SHEET, 2023), Pt (deposited via magnetron sputtering), Au and Cr (deposited via physical evaporation at 100 nm thickness). All samples were thoroughly polished prior to adhering the electrodes in order to exclude the presence of oxide skin. The electrical resistivity was determined with a Keithley 6487 device. Measurements were performed up to the limits of the apparatus, namely,: 1) loss of sensitivity at low voltage due to noise, 2) exceeding the current limit for high voltages (20 mA), 3) exceeding the maximum voltage (500 V).

At LNT, measurements were generally found to be compatible within ±25%, however deviations of up to 3 orders of magnitude appeared at room temperature (RT), as shown in Figure 5. In this latter case, Cr-electrodes yielded the lowest resistivity and a good ohmic behavior while the rest exhibited some form of charge blocking that was reduced as the field increased, approaching the Cr values. Therefore, the contact of the Cr-electrode was dimmed good across all temperatures explored in this work, and a systematic uncertainty assigned to the LNT measurements based on the dispersion between different electrode types.[1] Point-to-point variations of the resistivity were studied in different regions within the sample (near the four corners and at the center). Their relative spread, in the range 8%–30%, was used as an estimator of the sample uniformity.

Measurements as a function of temperature were taken in "drift" mode, i.e., by removing the sample from the LN bath and letting it gradually warm up, unimpeded. A PT-100 was epoxied directly to the sample (Figure 4-bottom). The first 5 s upon removal from the bath were excluded from the measurement to allow the sample to reach equilibrium, and the average of two identical series was taken. The combined uncertainty was estimated from their standard deviation, the point-to-point variations, and the electrode-to-electrode variations, added quadratically.

Last, measurements of the dielectric constant were performed, as its value impacts the signal induction process and the recovery time of resistive-protected detectors (e.g. (González Díaz et al., 2011; Riegler, 2016; Xie et al., 2021)). For practical reasons, a first assessment based on a standard LCR-meter with a limited range of 0.1 kHz–1 MHz was performed.

---

1 For few of the LNT series, and given that the low-frequency noise in the setup was not completely stable during the measurements, silver paint electrodes (thickness of few 100 μm) yielded a sensitivity down to lower voltages and were used instead.

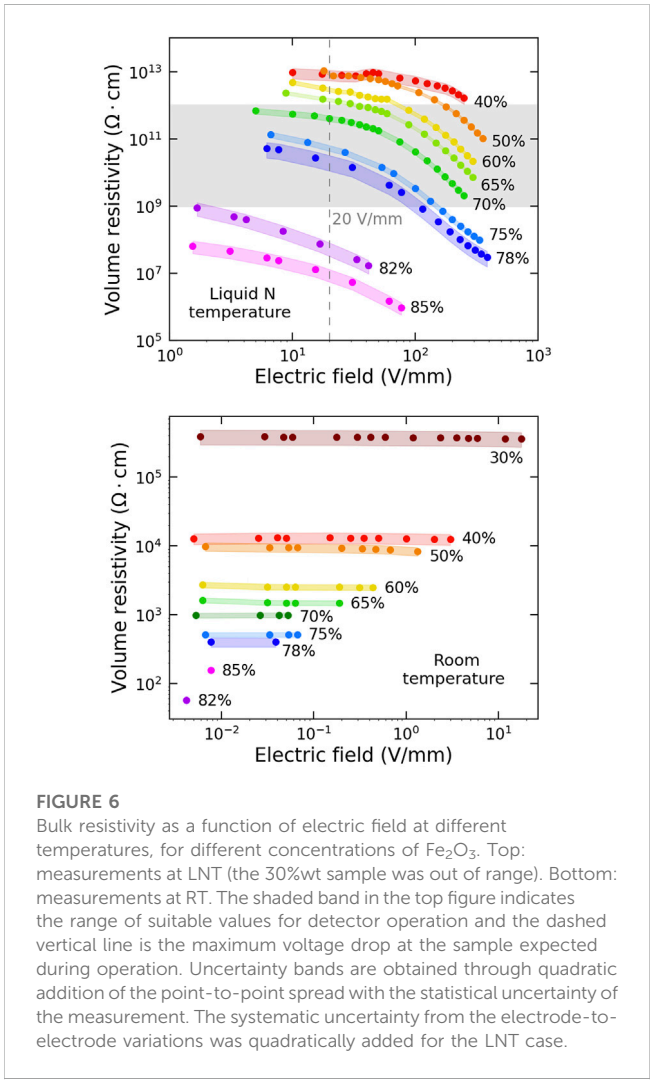

FIGURE 6
Bulk resistivity as a function of electric field at different temperatures, for different concentrations of $Fe_2O_3$. Top: measurements at LNT (the 30%wt sample was out of range). Bottom: measurements at RT. The shaded band in the top figure indicates the range of suitable values for detector operation and the dashed vertical line is the maximum voltage drop at the sample expected during operation. Uncertainty bands are obtained through quadratic addition of the point-to-point spread with the statistical uncertainty of the measurement. The systematic uncertainty from the electrode-to-electrode variations was quadratically added for the LNT case.

# 4 Measurements

## 4.1 Dependence with electric field

Resistivity measurements as a function of the external electric field for LNT and RT are given in Figure 6. All samples have thicknesses in the range 1–3 mm. A steep loss of linearity can be seen at LNT while no such effect could be observed at RT (albeit at much smaller fields, limited by the maximum current drawn by the picoammeter). During detector operation, a maximum (steady state) voltage drop of few 10′s of V across the ceramic sample can be anticipated, above which the gas amplification would start to deteriorate substantially. Hence, the 20 V value is highlighted with a vertical line for reference. A grey horizontal band, corresponding to values that are experimentally known to offer good spark quenching, was added following (Francke et al., 2003). It can be seen that, assuming a maximum voltage drop during operation in the order of few 10′s of V, samples in the $Fe_2O_3$ concentration range of 65%–78% wt would be well suited for operation close to LNT/LArT. For a low particle flux on the detector, resistivities as high as $10^{13}$ Ω·cm might be considered, potentially enabling the use of ceramics with concentrations down to 40%wt.

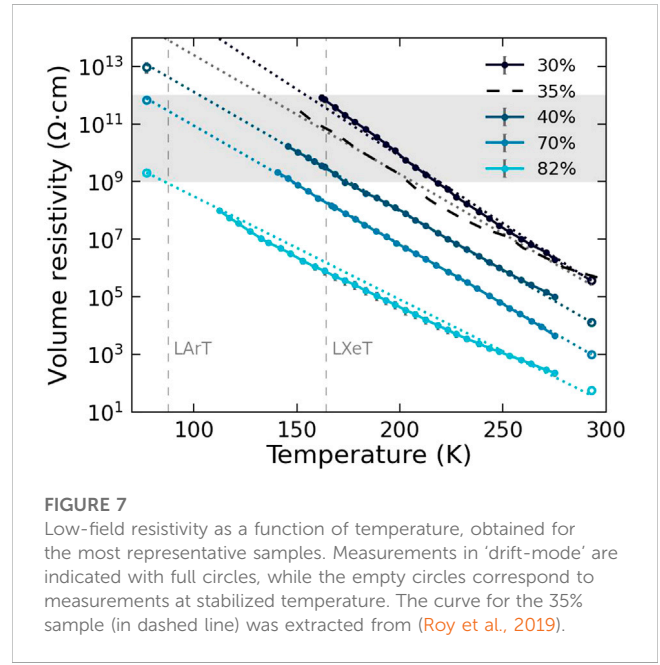

FIGURE 7
Low-field resistivity as a function of temperature, obtained for the most representative samples. Measurements in 'drift-mode' are indicated with full circles, while the empty circles correspond to measurements at stabilized temperature. The curve for the 35% sample (in dashed line) was extracted from (Roy et al., 2019).

The non-linearity observed in Figure 6-top can be expected to play a role during detector operation, as the aforementioned 20 V refers to steady-state (ss) conditions, while a streamer can cause local drops much higher than those, easily in the order of hundreds. The spark-quenching power depends on the entire evolution of the field in the amplification gas and through the plate during the discharge process. Hence it is reassuring to see that some of the samples stay within the grey zone up to much higher voltage drops than those expected in ss. Last, depending on the application, ss voltage drops well over 100 V can occur in some detectors operating at high event rates (see, e.g., (González-Díaz et al., 2006) for Multi-gap Resistive plate chambers -MRPCs), yet providing a usable response. The experimental range chosen in this work aims at covering this range of conditions.

In order to assess the electrical breakdown of the ceramics, several series of increasing and decreasing voltages in the range 0–5.5 kV were applied to both a 60% and a 75% sample at LNT through a CAEN471 power supply, and the current monitored. A monotonous behavior was observed up to 3 kV, above which the reproducibility of the *I-V* curve was lost. This voltage was then applied to the samples for several minutes, observing no instability. Hence, one can safely assume that the breakdown voltage is above 3 kV for mm-samples in the aforementioned concentration range, *a priori* suited for LNT/LArT operation. Strictly, for operation at around 3 kV voltage drop across the detector (as described later in Section 5), that value represents the maximum possible value of the *ss* voltage drop across the plate, under any circumstance (e.g., a self-sustained glow discharge in the gas).

## 4.2 Dependence with temperature

Resistivity measurements as a function of temperature are given in Figure 7 for the most representative samples. Results from a previous 35%wt-ceramic have been included as a dashed line (Roy et al., 2019), although a direct comparison is not possible given the less mature fabrication procedure and different post-processing. Measurements at stabilized temperature (LNT, RT) are included as open circles, and a weighted fit to an Arrhenius-like exponential law ($\log(\rho) = a - b \cdot T$) is shown as a dotted trend. The slopes are compiled in Table 1. The overall behavior supports the fact that, for practical purposes, good thermal uniformity is achieved in the sample during warm-up, and that the PT100 follows the sample temperature with good precision. Samples in the range 30–40%wt cross the shaded band at around the temperature of LXe (LXeT), as expected from (Roy et al., 2019).

TABLE 1 Parameters of the Arrhenius-like fit of the volume resistivity of $Fe_2O_3$/YSZ ceramic samples vs temperature: $\log(\rho) = a - b \cdot T$. Coefficient -a- is defined such that exp(a) has units of [$\Omega$ cm].

| Sample (%wt $Fe_2O_3$) | a | b · $10^{-2}$ ($K^{-1}$) |
|---|---|---|
| 82 | 12.08 ± 0.01 | 3.59 ± 0.01 |
| 70 | 15.07 ± 0.04 | 4.12 ± 0.02 |
| 40 | 16.27 ± 0.08 | 4.15 ± 0.03 |
| 35 | 17.54 ± 0.08 | 4.14 ± 0.03 |
| 30 | 19.20 ± 0.08 | 4.66 ± 0.03 |

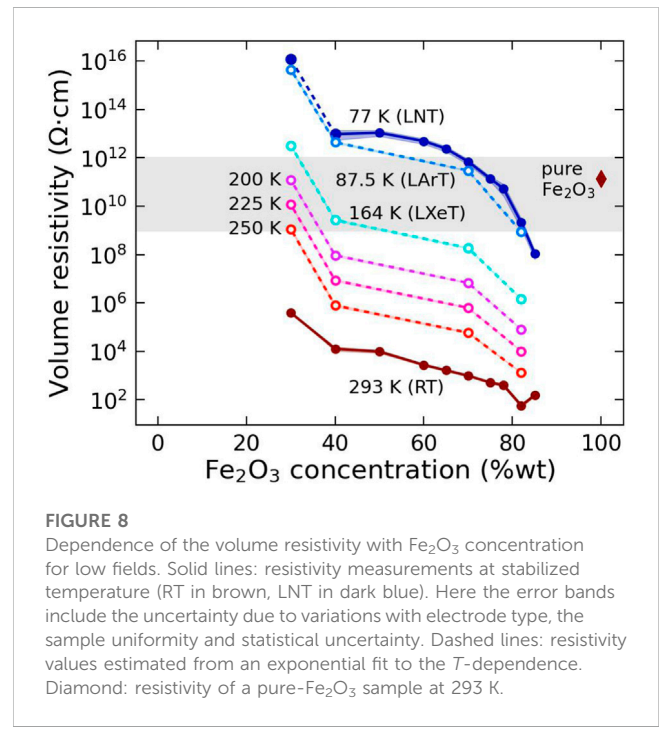

FIGURE 8
Dependence of the volume resistivity with $Fe_2O_3$ concentration for low fields. Solid lines: resistivity measurements at stabilized temperature (RT in brown, LNT in dark blue). Here the error bands include the uncertainty due to variations with electrode type, the sample uniformity and statistical uncertainty. Dashed lines: resistivity values estimated from an exponential fit to the $T$-dependence. Diamond: resistivity of a pure-$Fe_2O_3$ sample at 293 K.

Based on the observed temperature dependencies, it is possible to represent the low-field resistivities as a function of the concentration of $Fe_2O_3$, as done in Figure 8. Here, a reference measurement for pure $Fe_2O_3$ ceramics, performed at 293 K, has been included too (diamond). It is interesting to see that the trend as a function of $Fe_2O_3$ concentration is similar in the temperature range 77–293 K, except that it becomes steeper as the temperature decreases, in the extremes of the 30%–82%wt range. This becomes clearer when comparing the 77 K and 293 K series, for which this concentration range is more densely covered (open circles represent interpolations based on the fits to Figure 7). The reduction in electrical resistivity of $Fe_2O_3$ ceramics upon addition of ~15% YSZ amounts to a whopping nine-orders-of-magnitude gap. The resistivity measured for the 85%wt sample at RT provides a hint that such a concentration might be close to the inversion point.

## 4.3 Dependence with time/transported charge

The samples showed no strong dependence on humidity or time after storing them for about a year. Additional measurements were performed to indirectly assess the nature of the conductivity mechanism at room temperature and below, by applying voltage over time in a sustained manner. The situation is advantageous in that a very large charge can be transported at room temperature by profiting from the relatively large conductivity displayed in those conditions. Therefore, two sets of measurements were taken: 1) at LNT, for the maximum voltage of the picoammeter (500 V) and 2) at RT, for the maximum current (20 mA). Only the statistical uncertainty from the picoammeter was taken into account. The results are shown in Figure 9. At room temperature, no strong increase is seen up to 1 C/$cm^2$, a large number by high energy physics standards. Even if it is noticeable a 40% increase after about 4000 s, which might be attributed to the lack of stringent temperature control during these measurements, the sample clearly stabilizes after that. A measurement at LNT up to comparable charges should be carried out at a later stage, as it can not be confirmed at the moment that the conductivity mechanism is the same across temperatures. It should be noted, however, that for transported-charge densities well in excess of those anticipated by some future experiments (around 100 μC/$cm^2$ (Leardini et al., 2023)), no effect could be seen at LNT either, for the 70%wt sample (Figure 9-right). This result is encouraging, as the above conditions approach closely those for which detector operation was demonstrated (section 5).

## 4.4 Dependence with frequency

The transparency of a resistive plate to the signal induced by an avalanche in the gas depends on both its resistivity and capacitance. On the one hand, surface resistivities ($R_s$) above $10'$s of MΩ/sqr are known not to disturb signal induction at the time scale of avalanche formation (Riegler, 2016; Xie et al., 2021). For homogeneous materials such as the ones described here, where the relation $\rho/d = R_s$ may be assumed (d being the sample thickness), this condition is fulfilled by many orders of magnitude. On the other hand, if the coupling capacitance between the plate and pick-up electrode is too low, an additional loss can occur. According to the Schockley-Ramo theorem (Ramo, 1939), the current induced by an avalanche $I(t)$ under the hydrodynamic approximation is:

$$I(t) = q_e\ n(t)\ \mathbf{v}_d(t) \cdot \mathbf{E}_w \quad (1)$$

Here $q_e$ is the electron charge, $n(t)$ the number of charge carriers in the drift gap, $\mathbf{v}_d(t)$ their drift velocity and $\mathbf{E}_w$ the so called weighting field. Considering that for a parallel-plate geometry $v_d$ and $E_w$ are parallel, a 'coupling factor' might be defined as the ratio between the weighting field modulus in the protected structure and the unprotected one, as:

$$f = \frac{E_w|_{protected}}{E_w|_{unprotected}} \qquad (2)$$

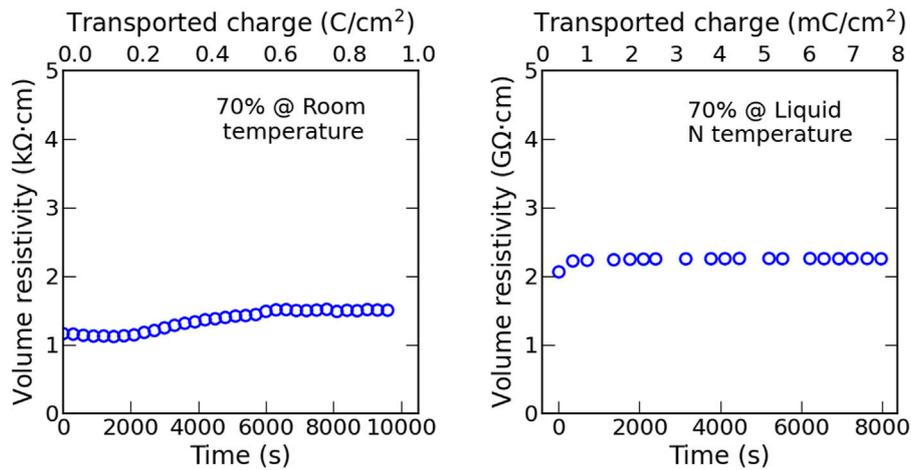

FIGURE 9
Stability of the electrical resistivity over time for a 70%wt Fe$_2$O$_3$ ceramics under sustained high voltage operation. The left figure (RT) was taken at the maximum current of the picoammeter. The right figure (LNT), at the maximum voltage (500 V).

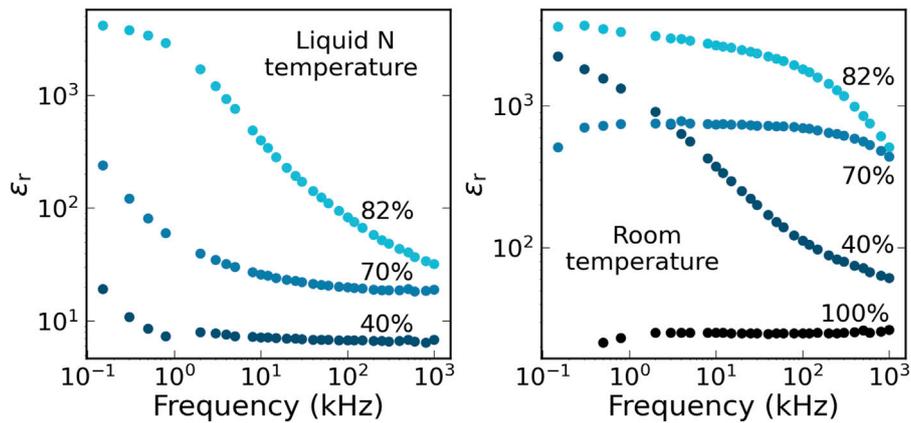

FIGURE 10
Dielectric constant ($\epsilon_r$) measured as a function of frequency at LNT (left) and RT (right), for selected samples.

$E_w$ is in this case nothing but the anode-cathode capacitance of the amplification structure, per unit area and divided by $\epsilon_0$ ($E_w = C_A/\epsilon_0$, e.g., (González-Díaz, 2012)). Specifically, for the RP-WELL/WELL geometries (Figure 11), the coupling factor evaluates to:

$$f = \frac{E_w|_{RP-WELL}}{E_w|_{WELL}} \simeq \frac{\epsilon_r}{\epsilon_r + d/g}, \qquad (3)$$

where $\epsilon_r$ is the dielectric constant of the plate, $g$ the depth of the gas holes and $d$ the plate thickness. In the limit when the plate has a very large dielectric constant or its thickness is much smaller than the hole depth, $f \rightarrow 1$.

Figure 10 shows the dielectric constant measured up to MHz, for illustrative purposes. If no additional relaxation mechanism would be present up to the GHz (avalanche) scale, the value of $\epsilon_r$ (1 MHz) might be used to evaluate Eq. 3, that we provide for illustration in Table 2. Although for insulators it is common to find relaxation at low frequencies (i.e., well below MHz), an additional elucidation of this aspect remains important.

Interestingly, these results point to a relatively high dielectric constant for operation close to LNT/LArT, as the 70%wt sample displays a value $\epsilon_r$ (1 MHz) $\simeq$ 20 in these conditions, about × 2-4 larger than other resistive materials that are usually found in resistive-plate detectors (González Díaz et al., 2011; Wang et al., 2013). Measurements for pure hematite at room temperature are found to be in the same range, in agreement with existing estimates (Lunt et al., 2013).

TABLE 2 Estimated coupling factor (defined in text) for RP-WELL structures of 0.8 mm-deep holes and 2 mm-thick resistive plates, at LNT.

| Sample (%wt $Fe_2O_3$) | $\epsilon_r$ | Coupling factor |
|---|---|---|
| 82 | 32.32 | 0.93 |
| 70 | 18.88 | 0.88 |
| 40 | 6.75 | 0.73 |

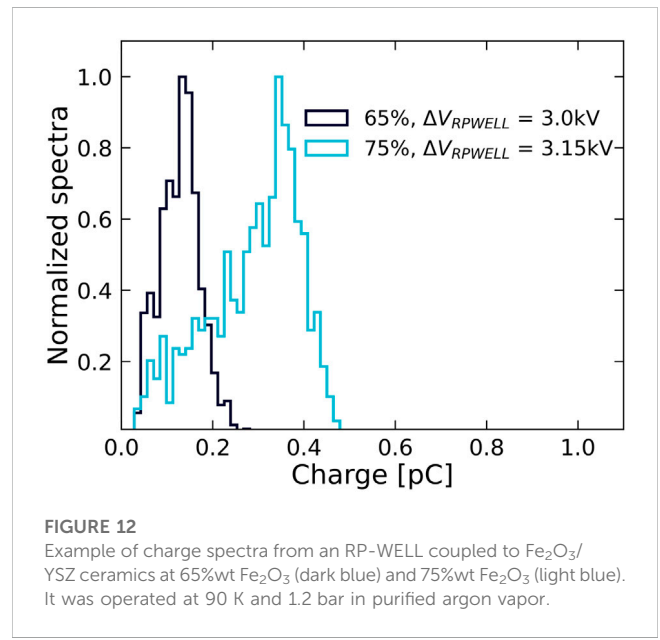

FIGURE 12
Example of charge spectra from an RP-WELL coupled to $Fe_2O_3$/YSZ ceramics at 65%wt $Fe_2O_3$ (dark blue) and 75%wt $Fe_2O_3$ (light blue). It was operated at 90 K and 1.2 bar in purified argon vapor.

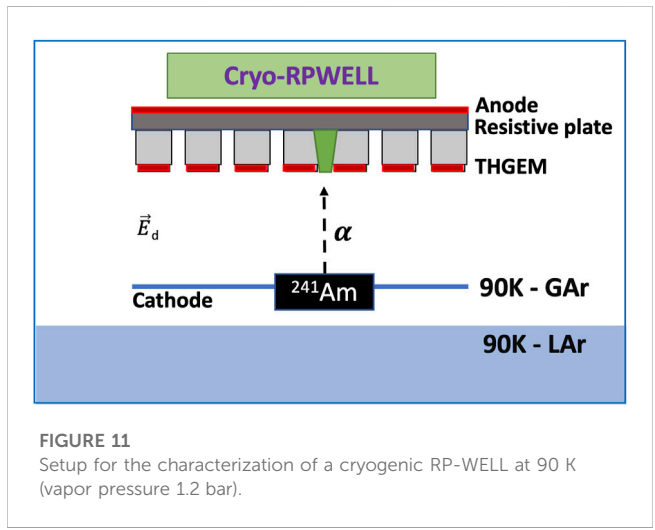

FIGURE 11
Setup for the characterization of a cryogenic RP-WELL at 90 K (vapor pressure 1.2 bar).

## 5 In-detector performance

Prior to this work, an RP-WELL was successfully operated at LXeT (163 K) under both X-rays and UV irradiation, reporting a superior performance compared to standard (unprotected) amplification (Roy et al., 2019). In that study, a ceramic plate made of $Fe_2O_3$/YSZ (35%wt $Fe_2O_3$) with a bulk resistivity of $\sim 10^{11}$ Ω·cm was used. Here we showcase the basic performance results from the recent operation of a cryogenic RP-WELL down to 90 K in purified argon vapor. The detector was assembled by means of a 0.8 mm-thick THGEM coupled to a pick-up anode via a ceramic plate (75%wt and 65%wt $Fe_2O_3$ were employed, corresponding to $\rho \sim 10^{10}$–$10^{11}$ Ω·cm at 90 K). Irradiation was performed with a collimated and attenuated $^{241}$Am alpha source installed on a metallic plate (cathode) 15 mm below the amplification structure. A drift/collection field of 0.5 kV/cm was set to the ionization region (Figure 11).

The source releases ionization electrons that are transferred to the RP-WELL holes and multiplied there. Typical charge-collected spectra for RP-WELL structures based on ceramic plates of 75%wt (light blue) and 65%wt (dark blue) are shown in Figure 12. They exhibit a full-energy peak corresponding to perpendicular emission, accompanied by a low-energy tail accounting for partial energy loss in the collimator at high angles. Given the energy of 4 MeV for the attenuated alpha particles in case of perpendicular incidence and the average energy to create an electron-ion pair, $W_I$ = 26.4 eV (Blum et al., 2008), the primary ionization in the gas can be estimated at 0.024 pC. Hence the charge spectra displayed in Figure 12 correspond to charge-multiplication factors in the range of 10. A more detailed analysis shows that the highest stable gain is 16, about × 3 the values obtained for an unprotected structure (details can be found in (Tesi et al., 2023)).

## 6 Discussion

Tunability of the resistivity of hematite ceramics by nine orders of magnitude, through addition of YSZ in different proportions, makes $Fe_2O_3$/YSZ ceramics suitable for gaseous-detector assemblies at temperatures from liquid nitrogen up to room temperature. Of particular interest are the 75%wt, 35%wt, and 100%wt samples, which show resistivities in the range $10^9$–$10^{12}$ Ω·cm at the LAr, LXe boiling temperatures and room temperature, respectively. The approximate agreement between the resistivity values of the new ceramic batch and an earlier version developed in (Roy et al., 2019) suggests that the overall characteristics of the final samples do not depend critically on the fabrication details.

Apart from the suitable resistivity range, the high breakdown voltage observed (above 3 kV) together with the strong hints of electronic conductivity and stability over time and transported charge, represent (a priori) essential assets for the intended application. Both literature and our measurements support the electron-conductivity hypothesis.

(i) On the low $Fe_2O_3$-concentration end (few %wt), Fe- and $Fe_2O_3$-doped YSZ ceramics have been extensively studied with concentrations up to around 5% mol and at high temperatures. Under these conditions, the conductivity is predominantly ionic. When the content of Fe-based impurities exceeds the solubility limit at around 6% mol (2.84%wt), paths of hematite have been reported to generate across the sample (Verkerk et al., 1982; Kharton, 1999; Boukamp, 2003).

(ii) On the medium $Fe_2O_3$-concentration region (20-50%wt), the aforementioned effect seems to induce electron movement at the Fe sites, either through disaggregated Fe ions at the grain-boundary (Boukamp, 2003),[2] or by electron hopping through Fe cations in the zirconia lattice (Slilaty, 1996; Kharton, 1999; Liao et al., 2011). At these unusual concentrations, mixed electronic-ionic conductivity has been reported at high temperature (Kharton, 1999; Guo, 2021), while the range below room temperature seems not to have been explored.

(iii) On the high $Fe_2O_3$-concentration end, below the YSZ percolation threshold, electron conductivity based on the polaron-hopping mechanism is known to take place (Liao et al., 2011; Ahart et al., 2022), and an enhancement in the conductivity is expected due to the higher amount of free electrons resulting from Zr donation (Liao et al., 2011).

Some (or all) of the above electron-conductivity mechanisms are expected to become dominant for low temperatures in case of medium and high $Fe_2O_3$-concentrations, given that the activation energy will be too low to enable ion conductivity (Kharton, 1999).

The existence of mixed electron-ion conductivity at room temperature is hinted in our data by indications of charge-blocking at low fields (Figure 5-bottom) as well as the non-monotonous behavior of the relaxation time with $Fe_2O_3$ concentration (Figure 10-right). While, based purely on the stability of the conductivity with transported charge, ion conductivity seems to be subdominant, additional indications can be found at LNT. In such conditions, charge blocking disappears for low electric fields and the relaxation times evolve monotonically with $Fe_2O_3$ concentration, facts that can be interpreted as a result of the conduction process becoming 'simpler' at low temperatures, and expectedly dominated by a form of electron conduction.

## 7 Conclusion

We demonstrate the maturity of $Fe_2O_3$/YSZ ceramic plates made in-house for use in gaseous detectors as resistive protection, across temperatures, specifically from 77 K (LNT) up to 300 K (RT). $Fe_2O_3$/YSZ ceramics are robust, display low-outgassing and their resistivity is tunable through adjustment of the $Fe_2O_3$ content. Optimal $Fe_2O_3$ concentrations are found to be 30-40%wt (LXeT), 65-75%wt (LArT) and 100%wt (RT). In the 65-75%wt range, for which 'in-detector' operation has been presented here for the first time at around LArT, the material has an approximately linear response up to 20 V voltage drop, and displays a breakdown voltage above 3 kV. Very good signal-coupling is expected by virtue of a relatively large dielectric constant around $\epsilon_r \sim 20$. No electrical-aging effect (e.g., exhaustion of charge carriers) was observed up to 8 mC/cm² (LNT) and 1 C/cm² (RT). According to these results, and existing literature, the dominant conductivity form of this family of ceramics is likely of electron origin, thus guaranteeing stability during run-time.

---

2   The possibility of the existence of a new phase at the grain-boundary has been advocated on that work too.

## Data availability statement

The raw data supporting the conclusion of this article will be made available by the authors, without undue reservation.

## Author contributions

LO-V performed the experimental characterisation of the ceramics and wrote sections of the manuscript; IP contributed substantially to the experimental characterisation of the ceramics; SL was responsible for overseeing the experimental activities, performing the initial characterisation of the ceramics, and detector operation; MM identified the ceramics to be built and provided the initial parameters for its fabrication; AC and RC were responsible for the ceramics manufacturing and process optimization; DG-D was responsible for funding procurement, project overseeing and writing; AT and LM were responsible for detector assembly and operation; CA and LC were responsible for funding procurement and studies related to detector assembly with materials of different thermal expansion coefficients; FG was the overall responsible for ceramics production and coordination. All authors contributed to the article and approved the submitted version.

## Acknowledgments

This research has been sponsored by RD51 funds through its "common project" initiative and has received financial support from Xunta de Galicia (Centro singular de investigación de Galicia accreditation 2019-2022), and by the "María de Maeztu" Units of Excellence program MDM-2016-0692. DGD was supported by the Ramón y Cajal program (Spain) under contract number RYC-2015-18820. We are thankful to the superconductivity group of USC (in particular to J. Mosqueira) for encouragement, discussions, and hardware support. We are also thankful to C. Moreno from the University of Cantabria for discussions on conductivity mechanisms and to A. Alegría from the Materials Physics Center of San Sebastián (CFM) for valuable feedback on the dielectric response measurements.

## Conflict of interest

The authors declare that the research was conducted in the absence of any commercial or financial relationships that could be construed as a potential conflict of interest.

The author(s) LC, CA, and LM declared that they were an editorial board member of Frontiers, at the time of submission. This had no impact on the peer review process and the final decision.

## Publisher's note